\documentclass{iopart}

\usepackage{iopams,setstack}
\usepackage{graphicx}
\usepackage{cite}
\usepackage{enumerate}

\newcommand{\bd}{\begin{displaymath}}
\newcommand{\ed}{\end{displaymath}}
\newcommand{\be}{\begin{equation}}
\newcommand{\ee}{\end{equation}}
\newcommand{\ba}{\begin{eqnarray}}
\newcommand{\ea}{\end{eqnarray}}

\begin{document}

\title[Boundary Bound Diffraction]{Boundary Bound Diffraction: A combined Spectral and Bohmian Analysis}

\author{Jalal Tounli, Aitor Alvarado and \'Angel S. Sanz}

\address{Department of Optics, Faculty of Physical Sciences,\\
Universidad Complutense de Madrid,\\
Pza.\ Ciencias 1, Ciudad Universitaria E-28040 Madrid, Spain}

\ead{\mailto{a.s.sanz@fis.ucm.es}}

\begin{abstract}
The diffraction-like process displayed by a spatially localized matter wave is here analyzed
in a case where the free evolution is frustrated by the presence of hard-wall-type boundaries
(beyond the initial localization region).
The phenomenon is investigated in the context of a nonrelativistic, spinless particle with mass
$m$ confined in a one-dimensional box, combining the spectral decomposition of the initially
localized wave function (treated as a coherent superposition of energy eigenfunctions) with a
dynamical analysis based on the hydrodynamic or Bohmian formulation of quantum mechanics.
Actually, such a decomposition has been used to devise a simple and efficient analytical
algorithm that simplifies the computation of velocity fields (flows) and trajectories.
As it is shown, the development of space-time patters inside the cavity
depends on three key elements: the shape of the initial wave function, the mass of the particle
considered, and the relative extension of the initial state with respect to the total length
spanned by the cavity.
From the spectral decomposition it is possible to identify how each one of these elements
contribute to the localized matter wave and its evolution; the Bohmian analysis, on the other
hand, reveals aspects connected to the diffraction dynamics and the subsequent appearance of
interference traits, particularly recurrences and full revivals of the initial state, which
constitute the source of the characteristic symmetries displayed by these patterns.
It is also found that, because of the presence of confining boundaries, even in cases of
increasingly large box lengths, no Fraunhofer-like diffraction features can be observed,
as happens when the same wave evolves in free space.
Although the analysis here is applied to matter waves, its methodology and conclusions are
also applicable to confined modes of electromagnetic radiation (e.g., light propagating
through optical fibers).
\end{abstract}


%03.65.Ta %Foundations of quantum mechanics; measurement theory
%03.75.Dg %Atom and neutron interferometry
%42.50.-p %Quantum optics
%42.50.Xa %Optical tests of quantum theory
%37.25.+k
%42.25.Hz %Interference

%PACS
%03.65.-w : Quantum mechanics (general)
%03.65.Ta : Foundations of quantum mechanics; measurement theory
%03.65.Xp : Tunneling, traversal time, quantum Zeno dynamics
%07.79.Cz : Scanning tunneling microscopes
%68.37.Ef : Scanning tunneling microscopy (including chemistry induced with STM)
%73.23.Hk : Coulomb blockade; single-electron tunneling
%82.20.Xr : Quantum effects in rate constants (tunneling, resonances, etc.)
%03.75.-b : Matter waves
%42.25.Hz : Interference

%%\submitto{\PS}

%\maketitle

%%%%%%%%%%%%%%%%%%%%%%%%%%%%%%%%%%%%%%%%%%%%%%%%%%%%%%%%%%%%%%%%%%%%%%%
%%%%%%%%%%%%%%%%%%%%%%%%%%%%%%%%%%%%%%%%%%%%%%%%%%%%%%%%%%%%%%%%%%%%%%%

\section{Introduction}
\label{sec1}

Localization, confinement and diffusivity settle a rather broad context, which includes myriads
of physical problems both fundamental and applied.
Pattern formation inside cavities constitutes, in this regard, a well-known topic in the literature.
Within this context, the present work is intended to render some light on the issue of how the presence
of boundaries affects the evolution (diffusion) of an initially localized matter wave, i.e., how boundaries
influence general diffraction (in time) processes.
Specifically, it is motivated by a need to elucidate whether there are universal key elements that determine the
evolution of a quantum carpet associated with a confined and initially localized quantum state, and how they
affect the corresponding quantum carpet that develops in space and time, thus providing a clue for new quantum
control mechanisms, for instance.
To that end, we have considered a rather simple, but illustrative problem, as it is the particle in a box,
where the system wave function at $t=0$ is a spatially localized state between two impenetrable walls,
i.e., a wave function with an everywhere vanishing amplitude except on a specific region between the box
boundaries, where it has finite values.
This state describes a nonrelativistic, spinless particle of mass $m$ in one dimension, with the
cavity length being $L$ and the localization region having a width $w$.
This can be the case, for example, of a neutron matter wave entering a waveguide \cite{calvo-waveguides-bk}
with rectangular cross section at a relatively high speed (compared to other transversal dynamical scales)
through a shutter characterized by an aperture smaller than the size of the incoming wave and with a particular
transmission function (not necessarily homogeneous along the opening).

Thus, to better appreciate the scope of the work, let us consider the case of localized matter waves.
As is well known, if such waves are spatially localized at a given time, they eventually undergo
delocalization once they are released and allowed to freely evolve.
This behavior is exhibited, for example, by trapped neutral atoms or condensates that are
suddenly released \cite{phillips:rmp:1998,grimm:aamop:2000}; once the atomic cloud is released,
it starts loosing localization as it spreads in time.
An analogous behavior is also displayed by matter wave crossing openings \cite{zeilinger:RMP:1988}
or gratings \cite{toennies:science:1994,toennies:PRL:1999,arndt:Nature:1999,arndt:NatCommun:2011}.
In this case, the passage through the opening or openings produces a number of transmitted (spatially
localized) beams that, with time, undergo delocalization by diffraction, giving rise to the appearance
in the long-time regime to a spatial redistribution of the probability --- the so-called Frauhofer
diffraction pattern.
The latter scenario is actually directly related to so-called diffraction in time phenomenon and the
Moshinsky shutter problem, introduced by Moshinsky in 1952 \cite{moshinsky:PhysRev:1952,moshinsky:AJP:1976}
to explain the appearance of transient terms in dynamical descriptions of resonance scattering
\cite{moshinsky:PhysRev:1951}.
In this phenomenon, diffraction-like features arise when a rather nonlocalized wave (e.g., a plane wave) is suddenly
truncated (by the action of a straight-edged shutter), thus producing a localized wave.
The subsequent transverse evolution of this wave is analogous to the evolution of a wave diffracted by
a real opening.
Diffraction in time has been a subject of interest in the literature ever since
\cite{kleber:PhysRep:1994,brukner:PRA:1997,schuch:JPA:2001,schuch:RevMexFis:2001,muga:JPA:2006,muga:PhysRep:2009,horwitz:FoundPhys:2007,goussev:PRA:2013},
being first confirmed with the time-domain analogous of single- and two-slit diffraction \cite{dalibard:PRL:1996}.
It is worth noting that, actually, this phenomenon is analogous to considering the evolution of a wave under paraxial
conditions, which allows to separate the problem into the longitudinal (fast) propagation, characterized by a classical-like
motion, and the transverse propagation, describable in terms of a Schr\"odinger equation of reduced dimensionality
\cite{sanz:JOSAA:2012,sanz:AOP:2015}.

The delocalization displayed by an initially localized matter wave can be, however, spatially limited
by adding some extra boundaries, which gives rise to additional phenomena.
Think, for instance, of such a matter wave as in the Moshinsky problem, i.e., and extended wave entering a cavity.
The initial localization of the ingoing state, produced by the size of the input shutter, evolves into
a rather symmetric patter in space and time displaying, at some positions and times, recurrences
and even full revivals of the initially localized state \cite{kaplan:PRA:2000,robinett:PhysRep:2004}.
Due to the similarity between these patterns and usual carpets, they are called {\it quantum carpets}
\cite{berry:PhysWorld:2001}, which may show fractal features under certain conditions \cite{berry:JPA:1996,sanz:JPA:2005}.
The emergence of such a pattern can be explained in terms of a complex interference process involving a number of energy
eigenstates of the confining cavity.
Now, although this is a bound effect, it is worth noting that an analogous situation also takes place
in the continuum when considering the transmission through a diffraction grating.
In such a case, a repetitive pattern arises by virtue of the so-called Talbot effect \cite{talbot:PhilosMag:1836,rayleigh:PhilosMag:1881},
used in optics, for instance, in lensless imaging applications \cite{wang:PRL:2017}, but in matter-wave interferometry
\cite{cronin:RMP:2009,arndt:OptExpress:2009} to probe the wave behavior and related properties of electrons \cite{cronin:NJP:2009},
atoms \cite{pritchard:PRA:1995} or large molecular systems \cite{arndt:PRL:2002,arndt:NatCommun:2011}.
Interestingly, in this case, although there are no physical
impenetrable walls constraining the evolution of the wave function, a series of periodic non-physical interference-mediated
walls arise as a consequence of the translational invariance symmetry displayed by the initial state \cite{sanz:JCP-Talbot:2007}.
These walls are such that the quantum flux confined within them remains the same all the way through (unless the translational
symmetry is spatially limited), generating patterns with recurrences and full revivals analogous to those observed in the problem
we are dealing with here.

At a more formal or fundamental level, if
we focus on the propagation itself of the matter wave inside the cavity, the above physical
situations turn into interesting diffusion-like problems,
where interference arises just because of the complex nature of
the probability amplitude or wave function, which is the quantity diffused across the
available configuration space (i.e., the cavity).
This brings out another interesting question, namely, that of the evolution of quantum
systems understood as a diffusion process.
In this regard, it is worth noting that, although Schr\"odinger's equation is commonly regarded
as a wave equation, formally speaking it has more in common with the heat-diffusion
equation,\footnote{Of course, there is an important limitation in this analogy: while the diffusion
equation arises from a conservation law, Schr\"odinger's equation does not describe the evolution
of any conserved quantity --- unlike the wave function, though, the associated probability density obeys
the continuity equation, which is a true transport equation, because it describes a conservation law.}
since both are parabolic partial differential equations \cite{morse-bk-2}.
That is, the evolution of matter waves and the propagation of heat both follow
an equation of the kind
\be
% \frac{\partial u}{\partial t} - D \frac{\partial^2 u}{\partial x^2} = 0 ,
 \frac{\partial u}{\partial t} - D \nabla^2 u = 0 ,
 \label{diff}
\ee
where the distribution or field variable $u({\bf r},t)$ specifies the state of the system
(i.e., the value of the probability amplitude or the temperature, respectively) at any
(allowed) position ${\bf r}$ at a given time $t$.
In this equation, the diffusivity constant $D$ plays a fundamental role, because it
determines the diffusion rate of $u$.
Now, while in the heat equation $D$ is a real number without a specific predetermined value,
when dealing with quantum systems, $D$ is a pure imaginary constant with value $i\hbar/2m$
\cite{furth:ZPhys:1933,comisar:PhysRev:1965} (other authors \cite{nelson:pr:1966} have
considered alternative approaches where $D$ is real, although its modulus remains the same).
On the one hand, by virtue of the complex-valuedness of $D$ we can observe interference
traits when dealing with quantum systems (which is not the case in heat transfer problems).
On the other hand, we find that, because $D$ depends on an external parameter, namely the mass
associated with the matter wave (in many-body problems there can be several of these constants,
each one with a different associated mass), the smaller the mass, the larger its diffusivity,
which translates into a faster spreading or delocalization rate.

In direct connection with the problem dealt with here, particularly the methodology that
is considered to tackle it, let us recall that is is a typical boundary value problem
described by Eq.~(\ref{diff}), where a given (initial) field function $u(x,0)$, specified
at $t=0$, is constrained to vanish at the boundaries at any time, i.e., $u(0,t) = u(L,t) =0$.
With these boundary conditions, the energy eigenfunctions or eigenmodes of the cavity form
a very convenient set of solutions for Eq.~(\ref{diff}).
As it is well-known from any elementary course on quantum mechanics, general solutions
to Eq.~(\ref{diff}) can be constructed by linearly combining such eigenfunctions.
So, initially localized solutions constitute a particular case, where their delocalization rate
is a function of $D$.
The subsequent evolution of these states can then be understood as a complex interference
process among all the involved eigenfunctions, but also analyzed in terms of the flux
associated with the wave function as a whole.
This means that the analysis is more efficient when combining the use of the energy spectrum
of the cavity, to analytically determine the evolution of the system quantum
state, with the numerically determined Bohmian trajectories, used to follow
the evolution of the system at a local level inside the box.
More specifically, in the problem here investigated it will be seen that
the former provides us with an efficient
method to compute the governing velocity (flux) fields (apart from other
quantities of interest, such as the probability density), while the latter
offers an insightful picture on how and why the probability evolves in the way
it does, explaining the pattern formation characteristic of these systems.
Besides its inherently fundamental interest to better understand the
processes of interference and recurrences in this kind of systems, as
it will be seen the analysis here conducted has an also intrinsic applied
interest as a ground for the development of efficient Bohmian-based
numerical methodologies.

Following the above prescriptions, it is shown that the three aspects ruling
the dynamical behavior of the system, which we are looking for, are: the shape of the initial
localized state, the particle mass, and the relative extension of the cavity with respect to the size
of the localization region of the state.
The first factor is related to the way how a shutter may transmit a matter wave incident on it.
From optics, we are used to homogeneous functions, although this may not be the case if there are
short-range interactions between the particles described by the matter wave (e.g., electrons, neutrons,
atoms, molecules, etc.) and the constituents of the material support where the shutter is, which are often
neglected, although they may have an important influence \cite{toennies:PRL:1999,toennies:PRL:2000,toennies:PRA:2000}.
The second factor, the mass of the particle, is important regarding the visualization of wave effects, since larger masses
should display classical-like features.
This introduces the question of the classical limit in a more natural way than the standard one typically based on
analyzing the behavior of energy eigenfunctions under some particular limit.
Finally, the third factor, the ratio between the size of the cavity and the size of the region where the state is localized
will render some light on the behavior exhibited by the system when it gets gradually freer (by free it has to be understood
the condition when the confining boundaries go to infinity).

The work has been organized as follows.
A general analysis of quantum diffraction in terms of eigenfunctions of the
infinite square well potential is presented in Section~\ref{sec2}, as well
as the method to compute the corresponding Bohmian trajectories in terms
of such eigenfunctions.
It is precisely by virtue of this analysis, where we readily notice that the
nonlinearity of the transport relation (Bohmian equation of motion or guidance
condition), that the mathematical superposition principle does not have a
direct physical counterpart.
Accordingly, the evolution of the quantum system cannot be naively
described by appealing to independent waves (eigenfunctions), since the
dynamics is governed by the collective effect of all of them as a whole.
This non-separability is a fundamental quantum trait coming from the
quantum phase, which translates into a non-crossing in the streamlines
or trajectories obtained in Bohmian mechanics.
In Section~\ref{sec3} this analysis is applied to the study of the three
different elements that influence the evolution of the quantum system
here considered: (1) the particular shape of the system initial state,
(2) its relative extension with respect to the size of the confining cavity, and
(3) the system mass.
These analyses will make use of a combination of quantum carpets and
associated sets of Bohmian trajectories; the first will provide us with an
overall perspective of the physics into play, while the latter will show us
the particularities of the probability flow.
To conclude, a series of remarks are summarized in Section~\ref{sec4}.

%%%%%%%%%%%%%%%%%%%%%%%%%%%%%%%%%%%%%%%%%%%%%%%%%%%%%%%%%%%%%%%%%%%%%%%%%%%%%%%%%%%
%%%%%%%%%%%%%%%%%%%%%%%%%%%%%%%%%%%%%%%%%%%%%%%%%%%%%%%%%%%%%%%%%%%%%%%%%%%%%%%%%%%

\section{Theory}
\label{sec2}

In the boundary value problem we are dealing with here, Eq.~(\ref{diff}) takes the form
of the time-dependent Schr\"odinger equation,
\be
 i\hbar\ \frac{\partial \psi(x,t)}{\partial t} =
  - \frac{\hbar^2}{2m}\frac{\partial^2}{\partial x^2}\ \! \psi(x,t) ,
 \label{eq2}
\ee
where $\psi(x,t)$ is constrained to the boundary condition $\psi(-L/2,t) = \psi(L/2,t) = 0$
at any time $t$ and $L$ is the total length of the box where the wave function is confined.
The initial condition is specified by the localized state
\be
 \psi_0(x) = \left\{ \begin{array}{lc}
   f(x) , \qquad & |x| \le w/2 \\
   0 , \qquad & w/2 < |x| \le L/2
  \end{array} \right. ,
 \label{eq36}
\ee
with $w$ being the effective size of the input shutter that allows the matter wave to
enter the cavity.

%%%%%%%%%%%%%%%%%%%%%%%%%%%%%%%%%%%%%%%%%%%%%%%%%%%%%%%%%%%%%%%%%%%%%%%

\subsection{Spectral analysis}
%\subsection{Setting up the boundary value problem}
\label{sec21}

At $t=0$, any general solution $\psi$ to (\ref{eq2}) can be recast in terms of a coherent superposition of
energy eigenfunctions, as
\be
 \psi_0(x) = \sum_\alpha c_\alpha \varphi_\alpha (x) .
 \label{eq20}
\ee
Since in one dimension the $\varphi_\alpha$ are real functions, the coefficients
are determined from the overlapping integral
\be
 c_\alpha = \int \varphi_\alpha(x) \psi_0(x) dx ,
 \label{eq21}
\ee
although the real-valuedness of $\varphi_\alpha$ does not ensure the real-valuedness of
$c_\alpha$, which also comes from the value of $\psi_0$ --- for instance, if $\psi_0(x)$
is a a traveling wave, e.g., a constant amplitude multiplied by a phase factor $e^{ikx}$,
then the $c_\alpha$ are complex-valued quantities.
In the particular case of the infinite square well here considered, the time-independent
eigenfunctions read as
\ba
 \varphi_\alpha^e(x) & = & \sqrt{\frac{2}{L}}\ \cos(k_\alpha x) ,
 \label{eq32} \\
 \varphi_\alpha^o(x) & = & \sqrt{\frac{2}{L}}\ \sin(k_\alpha x) ,
 \label{eq33}
\ea
with
\be
 k_\alpha= \frac{\pi \alpha}{L} .
 \label{eqkalp}
\ee
These solutions display, respectively, even ($e$) and odd ($o$) symmetry with respect to $x=0$,
i.e., $\phi_\alpha^e (-x) = \phi_\alpha^e (-x)$ for $\alpha = 2n - 1$ and
$\phi_\alpha^o (-x) = - \phi_\alpha^o (x)$ for $\alpha = 2n$, with $n = 1, 2, 3, \ldots$
Physically, these solutions indicate that only an integer number of
half wavelengths can be accommodated between the box boundaries, with
the largest half-wavelength being equal to the total distance, $L$,
between them.
The confining walls thus act in a way analogous to a space frequency (wavelength) filter,
removing any component that does not match such condition.

Following (\ref{eq20}), any general initial condition can then be recast as
\be
 \psi_0(x) = \sum_n c_{2n-1}^e \varphi_{2n-1}^e (x)
  + \sum_n c_{2n}^o \varphi_{2n}^o (x) .
 \label{eq35}
\ee
At any subsequent time, the wave function reads as
\ba
 \fl \psi(x,t) & = & \sum_\alpha c_\alpha \varphi_\alpha (x) e^{-iE_\alpha t/\hbar} \nonumber \\
 \fl & = & \sum_n c_{2n-1}^e \varphi_{2n-1}^e (x) e^{-i E_{2n-1} t/\hbar}
   + \sum_n c_{2n}^o \varphi_{2n}^o (x) e^{-i E_{2n} t/\hbar} ,
 \label{eq34}
\ea
since the time-evolution for $\varphi_\alpha$ is given by
\be
 \varphi_\alpha(x,t) = \varphi_\alpha(x) e^{-iE_\alpha t/\hbar} ,
 \label{tdph}
\ee
where
\be
 E_\alpha = \frac{p_\alpha^2}{2m}
% = \frac{\hbar^2 k_n^2}{2m}
     = \frac{\pi^2 \hbar^2 \alpha^2}{2mL^2}
 \label{eq38}
\ee
is the corresponding energy eigenvalue (with $p_\alpha = \hbar k_\alpha$).
Accordingly, if the transmitted wave function (initial condition) is described
by (\ref{eq36}), we find three possibilities:
\begin{itemize}
 \item[i)] If $f(x)$ is an even function, only the cosine series contributes to (\ref{eq34}).

 \item[ii)] If $f(x)$ is an odd function, only the sine series contributes to (\ref{eq34}).

 \item[iii)] If $f(x)$ has no definite parity (asymmetric function), a general combination of cosine
 and sine functions contributes to (\ref{eq34}).
\end{itemize}
In cases (i) and (ii) the parity or symmetry of the wave
function at any subsequent time is fully preserved.
The time-dependent phases (\ref{tdph}) only affect the amplitude of the
real and imaginary parts of the corresponding eigenfunctions, but not
their parity.
Hence, when the collective effect of all the contributing eigenfunctions
is taken into account, the parity of their total linear combination is
also preserved.
The same holds for $f(x) \in \mathbb{C}$.
In this case, the function can be split up into its real and imaginary
components, which are then recast in terms of the corresponding eigenfunction decompositions.
Contrary to directly operating over the full complex function, this procedure allows us
to take advantage of the symmetry of each component separately to perform the analysis.

Without loss of generality, from now on we shall consider the case of even-symmetric wave functions
with respect to $x=0$ (mirror symmetry), like (\ref{eq36}).
The corresponding time-dependent wave function reads as
\ba
 \psi(x,t) & = & \sqrt{\frac{2}{L}}
  \sum_n c_{2n-1} \cos(k_{2n-1} x)\ e^{-iE_{2n-1} t/\hbar}
 \nonumber \\
 & = & \sqrt{\frac{2}{L}}\ \! e^{-iE_1 t/\hbar}
  \sum_n c_{2n-1} \cos(k_{2n-1} x)\ e^{-i\omega_{2n-1,1} t} ,
%  = \sqrt{\frac{1}{2L}}
%  \sum_n c_{2n-1} e^{ik_{2n-1} x}\ e^{-iE_{2n-1} t/\hbar} ,
 \label{eq37}
\ea
with
\be
 \omega_{2n-1,1} \equiv \frac{E_{2n-1} - E_1}{\hbar}
  = \frac{2\pi^2 \hbar}{mL^2}\ \! (n - 1)n
 \label{eq40}
\ee
for $n \ge 2$ (for $n=1$, $\omega_{1,1}=0$).
From a dynamical viewpoint, the preceding global time-dependent phase factor
in (\ref{eq37}) can be neglected, as it is seen bellow with the aid of Bohmian
mechanics.
The behavior exhibited with time is ruled by the set of characteristic frequencies
$\omega_{2n-1,1}$, which introduce a series of related time-scales,
\be
 \tau_{2n-1,1} = \frac{2\pi}{\omega_{2n-1,1}}
 = \frac{mL^2}{\pi\hbar} \frac{1}{(n - 1)n} .
 \label{eq41}
\ee
Whenever the evolution time equals an integer multiple of the largest of these periods,
that is, the one for which $n = 2$,
\be
 \tau_{3,1} = \frac{mL^2}{2\pi\hbar} ,
 \label{eq42}
\ee
we observe a full recurrence of the wave function (leaving aside the aforementioned
global phase factor), since
\be
 \omega_{2n-1,1} \tau_{3,1} = (n-1)n\pi
\ee
is always an even integer of $\pi$.
From now on we shall refer to $\tau_{3,1}$ as the system {\it recurrence time}, which will
be denoted by $\tau_r$.
This is a {\it universal} quantity that does not depend on the initial shape of the wave
function or its width $w$, but only on the total length $L$ spanned by the cavity and the system
mass $m$.
Apart from $\tau_r$, there are other sub-multiples of this quantity for which fractional
recurrences can be observed, as will be seen in Sec.~\ref{sec3}.
In the particular case of initial wave functions characterized by non-differentiable
boundaries, the evolution is characterized by a series of alternate regular and fractal-like
(at irrational fractions of $\tau_r$) replicas \cite{berry:JPA:1996,sanz:JPA:2005}.
These systems present an additional symmetry known as selfsimilarity.

Apart from the time symmetry implicit in the fractional (or even fractal) recurrences
mentioned above, the time-evolving wave function also displays (spatial) mirror symmetry
(the symmetry of the initial state is preserved at any subsequent time) and time-reversal symmetry
with respect to $\tau_r$.
This can easily be seen through the probability density arising from (\ref{eq37}),
\ba
 \fl \rho(x,t) & = & \frac{2}{L} \sum_{n,n'} c_{2n-1} c_{2n'-1} \cos(k_{2n-1} x) \cos(k_{2n'-1} x)
       \cos(\omega_{2n-1,2n'-1} t) \nonumber \\
 \fl & = & \frac{2}{L} \sum_n c_n^2 \cos^2(k_{2n-1} x) \nonumber \\
% \fl & & + \frac{2}{L} \sum_{\substack{n,n'\\n \ne n'}}
% \fl & & + \frac{2}{L} \sum_{\case{n,n'}{n \ne n'}}
 \fl & & + \frac{2}{L} \sum_{n,n' \atop n \ne n'}
              c_{2n-1} c_{2n'-1} \cos(k_{2n-1} x) \cos(k_{2n'-1} x)
              \cos(\omega_{2n-1,2n'-1} t) ,
 \label{eq40bb}
\ea
where the bare sum of separate densities plus the sum of the coherence terms has been made
more apparent in the second equality.
Although the symmetries can be determined as well from (\ref{eq37}), the particular functional
form displayed by (\ref{eq40bb}) makes this expression more convenient to better understand the role
of the coherence terms in the appearance of revivals.
Actually, it is easy to see that, regardless of the complexity displayed by quantum carpets, if all
coherences are removed at once, we immediately recover $|\psi_0(x)|^2$, since the resulting $\rho(x,t)$
becomes independent of time.
This is of particular interest when introducing decoherence in this type of systems \cite{schleich:JMO:2000},
which can happen, for instance, by reproducing the experiment with entangled pairs of photon
\cite{song:PRL:2011,poem:PRL:2012}.

Thus, getting back to (\ref{eq40bb}), in order to determine other symmetry structures, observe that
we have a large set of frequencies, which follow the general expression
\be
 \fl \omega_{2n-1,2n'-1} = \frac{E_{2n-1} - E_{2n'-1}}{\hbar}
  = \frac{2\pi^2\hbar^2}{mL^2}\ \big[ (2n-1)n - (2n'-1) n' \big] .
 \label{eq40nn}
\ee
This expression generalizes the previous expression (\ref{eq40}) to any pair of $n$ and $n'$ components
[in (\ref{eq40}), we had $n'=1$].
If we exchange $x$ by $-x$ in Eq.~(\ref{eq40bb}), we readily find that
\be
 \rho(-x,t) = \rho(x,t) ,
\ee
which is satisfied at any time $t$.
According to this symmetry, all what happens in one half of the space inside
the box has a mirror replica in the other half.

Regarding the time symmetry, consider two times symmetrically picked up around
the recurrence time, i.e., $t_1 = \tau_r/2 - t$ and $t_2 = \tau_r/2 + t$.
Evaluating (\ref{eq40bb}) at $t_1$ and then at $t_2$, we find
\be
 \rho(x,t_1) = \rho(x,t_2) ,
\ee
with $\tau_r/2$ playing the role of a critical or inversion time.
This means that the density evolves in time developing a series of interference
features until $t = \tau_r/2$; then, it undergoes an {\it involution}, passing through
all the previous stages until it eventually {\it recollapses}, reaching a state equal
to the departure state (with respect to the probability density, since the wave function,
as seen above, accumulates a global phase factor that makes it to be exactly the same we
had at the beginning).
To some extent the situation is analogous to a closed universe (in terms of density
rather than shape), where after reaching maximum expansion, it collapses again.
This behavior is independent of the shape of the initial state, the system
mass, or the extension of the confining box; in all cases an expansion
(diffraction) and recollapse of the system is expected (unless dissipation and/or
decoherence are somehow present).

%%%%%%%%%%%%%%%%%%%%%%%%%%%%%%%%%%%%%%%%%%%%%%%%%%%%%%%%%%%%%%%%%%%%%%%

\subsection{Bohmian analysis}
\label{sec23}

The above spectral analysis allows us to understand the evolution of the matter
wave inside the cavity by means of a complex interference process among different
energy eigenfunctions.
Instead of appealing to the energy representation, the same process can also be understood
in the configuration representation, where the spatial interference observed (see Sec.~\ref{sec3})
is usually explained in terms of semi-classical argumentations based on the computation of classical
orbits \cite{robinett:AJP:2000}.
In this regard, Bohmian mechanics provides us with an alternative and complementary analysis tool based
on locally monitoring the quantum flux with the aid of trajectories.
This is possible by means of the nonlinear (polar) transformation
\begin{equation}
 \psi ({\bf r},t) =
  \rho^{1/2}({\bf r},t) \ \! {\rm e}^{{\rm i} S({\bf r},t)/\hbar} ,
 \label{eq50}
\end{equation}
which recast a complex-valued field ($\psi$) in terms of two real-valued fields, namely the
probability density, $\rho$, and the (wave function) phase, $S$.
After substitution of (\ref{eq50}) into the time-dependent
Schr\"odinger equation, two (real-valued) coupled partial differential
equations arise,
\ba
 \frac{\partial \rho}{\partial t} & + & \nabla \cdot {\bf J} = 0 ,
 \label{eq51} \\
 \frac{\partial S}{\partial t} & + &
  \frac{(\nabla S)^2}{2m} + V + Q = 0 ,
 \label{eq52}
\end{eqnarray}
with ${\bf J} = \rho \nabla S/m$ being the usual probability current or
quantum flux \cite{schiff-bk},
\begin{equation}
 {\bf J} = D \left( \psi \nabla \psi^* - \psi^* \nabla \psi \right) .
% = D \psi \overset{\leftrightarrow}{\nabla} \psi^*
 \label{eq54bb}
\end{equation}
Equation (\ref{eq51}) is the well-known continuity equation for the conservation of
the probability, while (\ref{eq52}), more interesting from a dynamical viewpoint,
is the quantum Hamilton-Jacobi equation governing the particle motion under
the action of a total effective potential: $V_{\rm eff} = V + Q$.
The last term in the left-hand side of (\ref{eq52}),
\begin{equation}
 Q = - \frac{\hbar^2}{2m}\frac{\nabla^2 \rho^{1/2}}{\rho^{1/2}} ,
 \label{eq53}
\end{equation}
is the so-called {\it quantum potential}, which depends on the system quantum
state through the density field.

In the classical Hamilton-Jacobi theory, $S$ represents the mechanical action
of the system at a time $t$, with the trajectories describing the system
evolution corresponding to the paths perpendicular to the
constant-action surfaces at each time.
Given that quantum mechanics is just a wave theory (regardless of the
physical meaning that one may wish to assign to the wave function),
one can proceed similarly according to the above polar transformation,
which allows us to operate with $S$ in analogy to its classical counterpart.
Accordingly, the classical-like concept of trajectory emerges in
quantum mechanics in a natural fashion: particle trajectories are defined
as the solutions of an equation of motion that admits different functional
(convenient) functional forms,
\begin{equation}
 \dot{\bf r} = \frac{\nabla S}{m} = \frac{\bf J}{\rho}
  = \frac{\hbar}{m} \ {\rm Im} \left\{ \psi^{-1} \nabla \psi \right\}
  = \frac{1}{m} \ {\rm Re} \left\{ \frac{\hat{p}\psi}{\psi} \right\} .
 \label{eq54}
\end{equation}
Here, $\hat{p} = - i \hbar \nabla$ is the usual momentum operator in
the configuration representation.
Notice that ${\bf v} = \dot{\bf r}$ specifies a velocity field
predetermined at each time by the value of the system wave function
$\psi$ (through its phase $S$ or, equivalently, the flux ${\bf J}$).
This is particularly interesting at $t=0$, since the initial momentum
is predetermined by the initial wave function, $\psi_0$.
This means that trajectories (or, equivalently, quantum fluxes) must
evolve in time following a given prescription, this being a dynamical
manifestation of the so-called {\it quantum coherence}.
Notice here the difference with respect to point-like classical
mechanical systems, with their initial momenta being independent
of their initial positions.
In this sense, although both quantum and classical systems evolve
under a similar equation, namely a Hamilton-Jacobi equation, they
cannot be directly compared because their dynamics are very different.
In the quantum (Bohmian) case, dynamics take place in
configuration space, thus only being dependent on coordinates
[momenta are fixed at each point by the phase field, as it can be
inferred from Eq.~(\ref{eq54})], while classical dynamics develop
in phase space, where coordinates and momenta are both
independent variables.

Taking into account the explicit functional form of the wave function (\ref{eq37}),
the equation of motion (\ref{eq54}) for a general superposition of energy eigenfunctions
takes the form
\ba
 \dot{x} = \frac{1}{m}
 \frac{\sum_{\alpha,\alpha'} |c_\alpha| |c_{\alpha'}| \sin(k_\alpha x) \cos(k_{\alpha'} x)
       \sin (\omega_{\alpha,\alpha'} t + \delta_{\alpha,\alpha'})}
      {\sum_{\alpha,\alpha'} |c_\alpha| |c_{\alpha'}| \cos(k_\alpha x) \cos(k_{\alpha'} x)
       \cos(\omega_{\alpha,\alpha'} t + \delta_{\alpha,\alpha'})} .
 \label{eq55}
\ea
In this equation, the coefficients preceding each eigenfunction have been recast in polar form,
\be
 c_\alpha = |c_\alpha| e^{i\delta_\alpha} ,
\ee
assuming that they may also introduce a complex phase factor.
This explains the phase shifts $\delta_{\alpha,\alpha'} = \delta_\alpha - \delta_{\alpha'}$ that
appear in both the numerator and the denominator of this equation.
In our particular case, though, the $c_\alpha$ coefficients are real, for which
$\delta_\alpha = 0$, and therefore Eq.~(\ref{eq55}) acquires the simpler functional form
\ba
 \dot{x} = \frac{1}{m}
 \frac{\sum_{\alpha,\alpha'} c_\alpha c_{\alpha'} \sin(k_\alpha x) \cos(k_{\alpha'} x)
       \sin (\omega_{\alpha,\alpha'} t)}
      {\sum_{\alpha,\alpha'} c_\alpha c_{\alpha'} \cos(k_\alpha x) \cos(k_{\alpha'} x)
       \cos(\omega_{\alpha,\alpha'} t)}  .
 \label{eq55b}
\ea

Because the velocity field (\ref{eq55b}) satisfies exactly the same
symmetry conditions displayed by $\psi_0$,
the trajectories will also manifest this kind of overall feature
\cite{goldstein:PRE:1999}.
However, it is also possible to go the other way around and extract
valuable information about the topology displayed by the trajectories
and, from it, about the dynamical behavior of the system.
For example, the fact that the solution trajectories obtained from
(\ref{eq55b}) cannot cross the same space point at the same time
\cite{sanz:JPA:2008,sanz:JCP-Talbot:2007} implies that the dynamical behavior of the system
can be split up into different domains.
Specifically, in the cases considered here the mirror symmetry
with respect to $x=0$ translates into two separate dynamical regions,
with the trajectories from one domain never penetrating the other one,
and vice versa.
This can easily be inferred from the fact that $v(x=0) = 0$ at any time,
which means that the quantum flux splits up into two separate fluxes,
each one confined in one half of the box \cite{sanz:JPA:2008,sanz:JCP-Talbot:2007}.

%%%%%%%%%%%%%%%%%%%%%%%%%%%%%%%%%%%%%%%%%%%%%%%%%%%%%%%%%%%%%%%%%%%%%%%%%%%%

\section{Numerical simulations}
\label{sec3}

Diffraction is typically associated with functions or states characterized by well-defined edges, even if this implies their
non-differentiability on some particular space points.
This is an interesting aspect to be analyzed, for such edges strongly determine not only the speed or rate of the system
diffusion inside the box, but also the type of recurrences that can be expected, which determines in the last instance the
transport properties of the box if it simulates, for instance, an optical fiber or the depth of a slit to diffract matter waves
(e.g., electrons, neutrons or atoms).
Physically, this initial shape can be related to the transmission function associated with the shutter, which does not necessarily
has to be a flat function all over its extension.
On the contrary, it can be given in terms of a modulation function, as happens when we insert optical filters for light or, in the
case of matter waves, specified by the effect of the potential mediating the interaction between the diffracted particle and the
constituents of the aperture.
Analogously, as can be noticed through the functional form displayed by the time-dependent wave function (\ref{eq37})
and its recurrence time (\ref{eq41}) [or its frequency (\ref{eq40})], the system mass $m$ as well as the box length $L$
(in relation to the dimension of the shutter or, equivalently, the space region where the initial wave function is nonzero)
are also going to play an important role in the subsequent evolution of the wave and the type of interference features that
it will develop with time.
Below, the effects of these three aspects on the system diffusion are going to be discussed with the aid of a series of
numerical simulations based on the analytical forms (\ref{eq40bb}) for the probability densities, and (\ref{eq55b}) for the
velocity field and, by
integration, the corresponding Bohmian trajectories.

To better understand the effects of these influential aspects and particulary to acquire a more
quantitative idea of them, in the analysis we are going to consider some quantities of interest.
One of them is the {\it overlapping probability}, here defined as the overlap between the exact
wave function, $\Psi(x,t)$, and its associated series truncated at the $N$th term, $\Psi_N(x,t)$,
i.e.,
\be
 P_N = \int \Psi_N^*(x,t)  \Psi(x,t) dx = \sum_{n=1}^N c_{2n-1}^2 .
 \label{pntest}
\ee
This quantity provides us with a direct measure of the convergence of the series and, therefore,
how the above parameters influence the superposition and the subsequent interference traits
developed along time.
Another two quantities of interest are the {\it relative weight} $c_{2n-1}^2$ of each component
of the superposition as a function of $n$, just to get an ide on the relevance of each contributing
eigenfunction, and the {\it expectation value of the energy} for $\Psi_N(x,t)$,
\be
 \langle H \rangle_N =
  \frac{\sum_{n=1}^N |c_{2n-1}|^2 E_{2n-1}}{P_N} ,
 \label{enertest}
\ee
which is also a measure of convergence in terms of the energy added by each component to the superposition.

%%%%%%%%%%%%%%%%%%%%%%%%%%%%%%%%%%%%%%%%%%%%%%%%%%%%%%%%%%%%%%%%%%%%%%%

\subsection{Influence of the shape: matter-wave diffraction}
\label{sec31}

\begin{table}[t]
 \centering
 \begin{tabular}{|c|c|c|}
  \hline \hline
   Shape of $\psi_0$ & $f(x)$ & $c_\alpha\quad$ $(\alpha=2n-1)$ \\
  \hline \hline
   Square & $\displaystyle \frac{1}{\sqrt{w}}$ & $\displaystyle \sqrt{\frac{2w}{L}}\ {\rm sinc} \left( k_\alpha w/2 \right)$ \\ \hline
   Triangle & $\displaystyle \sqrt{\frac{3}{w}} \left( 1 - \frac{2|x|}{w} \right)$ & $\displaystyle \sqrt{\frac{3w}{2L}}\ {\rm sinc}^2 \left( k_\alpha w/4 \right)$ \\ \hline
  Parabola & $\displaystyle \sqrt{\frac{15}{8w}} \left[ 1 - \left( \frac{2x}{w} \right)^2 \right]$ &
              $\displaystyle \frac{4}{w}\sqrt{\frac{15}{wL}}\frac{1}{k_\alpha^2} \big[ {\rm sinc} \left( k_\alpha w/2 \right) - \cos \left( k_\alpha w/2 \right) \big]$ \\ \hline
  Half-cosine &
              \parbox{4.5cm}{$\displaystyle \sqrt{\frac{2}{w}}\ \cos \left( k_0 x \right),\quad k_0 = \pi/w$} &
              \parbox{7cm}{$\displaystyle \frac{4}{\sqrt{Lw}} \left( \frac{k_0}{k_0^2 - k_\alpha^2} \right) \cos \left( k_\alpha w/2 \right)$,\quad for $k_\alpha \ne k_0$ \\  \\
              $\displaystyle \sqrt{\frac{w}{L}}$, \qquad \qquad \qquad \qquad \quad for $k_\alpha = k_0$} \\ \hline
  Half-cosine squared &
              \parbox{4.5cm}{$\displaystyle \sqrt{\frac{8}{3w}}\ \cos^2 \left( k_0 x \right),\ \ k_0 = \pi/w$} &
              \parbox{7.5cm}{$\displaystyle \sqrt{\frac{4w}{3L}} \left[ \frac{(2k_0)^2}{(2k_0)^2 - k_\alpha^2} \right] {\rm sinc} \left( k_\alpha w/2 \right)$,\quad for $k_\alpha \ne k_0$ \\  \\
              $\displaystyle \sqrt{\frac{1}{3Lw}} \frac{16}{3k_0}$, \quad \qquad \qquad \qquad \quad for $k_\alpha = k_0$} \\ \hline
   Gaussian & $\displaystyle \left( \frac{1}{2\pi\sigma_0^2} \right)^{1/4} e^{-x^2/4\sigma_0^2}$ &
              $\displaystyle \sqrt{\frac{2}{L}} \left( 8\pi\sigma_0^2 \right)^{1/4} e^{-\sigma_0^2 k_\alpha^2}$ \\ \hline \hline
 \end{tabular}
 \caption{\label{tab1} Different functional shapes considered in this
  work for the (simulated) diffracted wave function $\psi_0$ and their
  corresponding Fourier components.
  The width of the Gaussian function, $\sigma_0=w/2\pi$, has been chosen so that, in first
  approximation, it equals the squared half-cosine function (moreover, with this value, the
  corresponding probability density has decreased to about $4\%$ at $|x| = w/2$).
  These wave functions are displayed in Fig.~\ref{fig1}(a).}
  \vspace{0.2cm}
\end{table}

\begin{figure}[t]
 \centering
 \includegraphics[width=\textwidth]{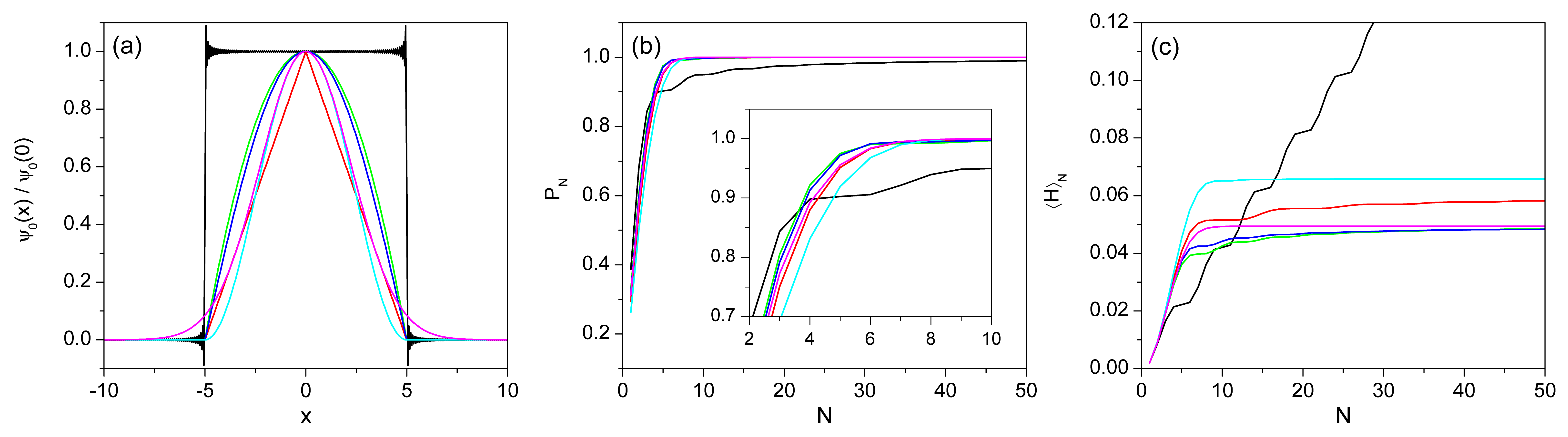}
 \caption{\label{fig1}
  (a) Reconstruction of the six wave functions described in Table~\ref{tab1}
  with a total of $N=500$ eigenfunctions each: square (black), triangle (red), parabola (green),
  half cosine (blue), half-cosine squared (cyan), and Gaussian (magenta).
  For a better comparison, all wave functions have been normalized to their
  value at $x=0$ ($\Psi_0(0)$).
  In all cases: $L=50$, $w=10$, and $\hbar = m = 1$.
  (b) Probability $P_N$ (\ref{pntest}) as a function of the number $N$ of contributing eigenfunctions.
  For visual clarity, an enlargement of $P_N$ for low $N$ is shown in the inset.
  (c) Expectation value of the Hamiltonian (\ref{enertest}) as a function
  of the number $N$ of eigenfunctions.}
\end{figure}

First we are going to analyze the spreading or diffusion and subsequent interference and recurrences in the position (configuration) space of
a series of diffracted waves $\psi_0$, an analysis that emphasizes the relationship between such traits or phenomena and the relative curvature
of the diffracted function.
We have chosen a series of functional forms $f(x)$ for the initially localized wave packet (see Table~\ref{tab1}) in a range that
covers various intermediate functions, from the square function to the Gaussian wave packet [see Fig.~\ref{fig1}(a)].
The square function constitutes a paradigm of transmission function in both optics and quantum mechanics, although more
realistic in the first case than in the latter due to the faster propagation of light with respect to usual matter waves.
The Gaussian function, in many cases of physical interest, has a more convenient computationally functional
form, apart from being more realistic when short-range interactions with the opening boundaries are non-negligible.
As intermediate cases we have chosen a triangle, a parabola, a half-cosine and a half-cosine squared, which present
different degrees of differentiability and curvature.
All these intermediate cases have been chosen in such a way that they vanish at $x = \pm w/2$.
As for the Gaussian function, its width has been chosen in a way that, in first approximation, its functional form equals that of the half-cosine
squared, as can be seen in Fig.~\ref{fig1}(a) by means of the overlap of both functions for $|x| \lesssim 1/3$.
As can be seen in the figure, the triangular function is quite close to these two functions, while the parabola and half-cosine functions are closer
between themselves.
The decomposition of all these functions in terms of energy eigenfunctions of the infinite square well
potential can be seen in Table~\ref{tab1} in terms of the generic $\alpha$th coefficient, with $\alpha = 2n-1$.
All the initial ans\"atze considered in Table~\ref{tab1} share a general common feature worth mentioning, as can be
noticed in their eigenfunction decomposition: $\Psi_0$ does not depend on the system mass ($m$), but on the ratio $w/L$.
However, despite this fact, the dynamics displayed by $\Psi_0$ is strongly dependent on the mass, since it
appears in a key dynamical element, namely the frequencies (\ref{eq40}) and (\ref{eq40nn}).
According to the expressions for these frequencies, the larger the mass, the lower the frequency (energy).
These facts are related to the time-reverse and mirror symmetries above discussed.
Large masses and/or box lengths will imply longer recurrence times, i.e.,
slower dynamics, as will be seen in more detail in Sec.~\ref{sec32}.

Let us now discuss in more detail some properties of the six wave functions.
In Fig.~\ref{fig1}(a) we observe a reconstruction of all the functions
considered.
A total of 500 eigenfunctions has been considered in each case.
As it can be seen (and is well known), the shape of each function is
well converged, except the square wave function due to the discontinuity
at $x = \pm w/2$.
This mismatch is produced by the well-known Wilbraham-Gibbs phenomenon
\cite{wilbraham:CDMJ:1848,gibbs:Nature:1898,gibbs:Nature:1899},
which, in the context of Fourier analysis, states that a Fourier series will display a finite increase or
decrease of the value of the sampled function at those points where the function has a
discontinuity, independently of how many Fourier components are considered in the series.
Strictly speaking, although we are not performing Fourier analysis, the decomposition of the
function in terms of a basis set associated with a certain potential function (an infinite square
potential) is analogous.
As a consequence, although the sum of components approaches very slowly the normalization to unity, as
seen in Fig.~\ref{fig1}(b), the expectation value of the Hamiltonian is unbound, as shown in Fig.~\ref{fig1}(c).
This is connected to fractal like features in the evolution of the square
function \cite{berry:JPA:1996} each time that one looks at a time that
is an irrational submultiple of $\tau_r$.
This behavior quickly disappears as the discontinuity at $\pm w/2$ also disappears,
as can be seen in the other cases shown in Figs.~\ref{fig1}(b) and (c).
In Fig.~\ref{fig1}(b) the inset shows how $P_N$ approaches the unity with a very few eigenfunctions
for all functions; in Fig.~\ref{fig1}(c) it is shown that, in spite of the nondifferentiability at $\pm w/2$ for the
triangle, the parabola and the half cosine, the expectation value of their energies remain bound.
In this regard, because of the nondifferentiability also at $x=0$, the convergence of the energy for the triangle
function is relatively slower than the other cases, since the number of contributing eigenfunctions is larger.

\begin{figure}
 \centering
 \includegraphics[width=\textwidth]{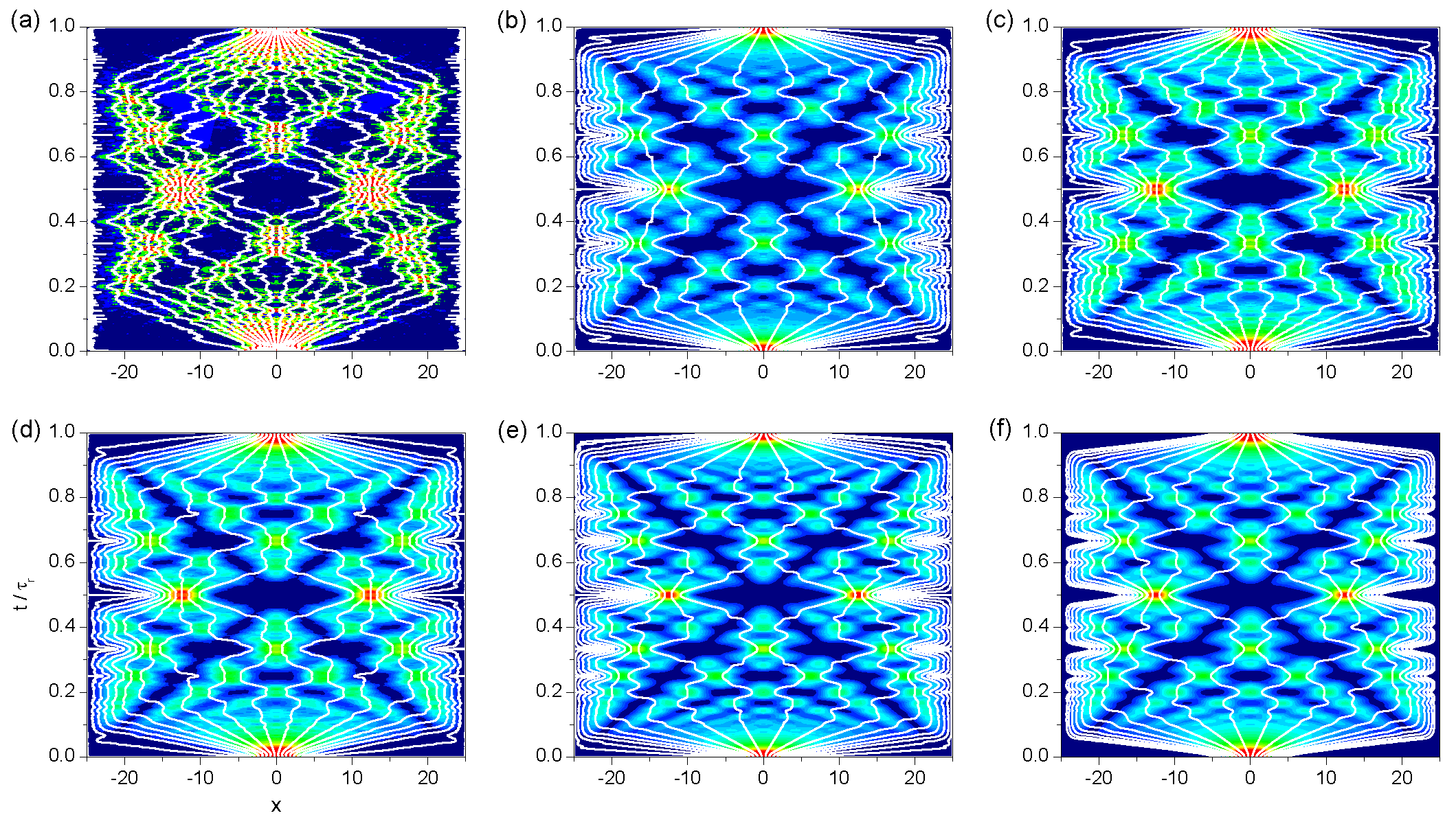}
 \includegraphics[width=\textwidth]{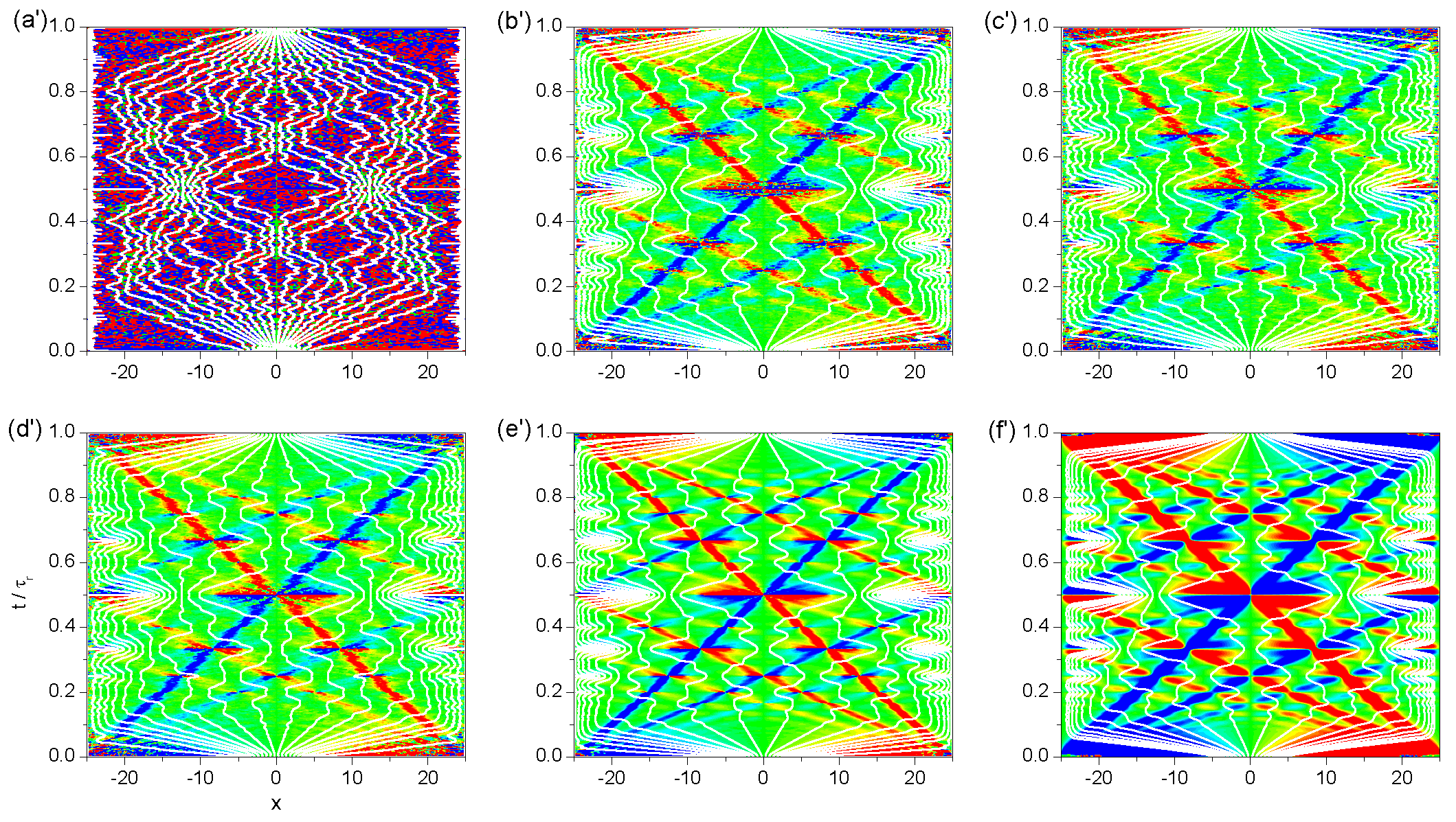}
 \caption{\label{fig2}
  Contour-plots showing the quantum carpets displayed by the six wave
  functions described in Table~\ref{tab1} along their evolution:
  (a) square, (b) triangle, (c) parabola, (d) cosine, (e) cosine square,
  and (f) Gaussian.
  The six upper panels represent the probability density, while the six
  lower panels (labeled with primes) refer to the corresponding velocity
  fields.
  Sets of quantum trajectories have been superimposed in order to
  illustrate the dynamical evolution of the flux in each case.
  In all cases: $N=500$, $L=50$, $w=10$, and $\hbar = m = 1$.
  For visual clarity, the probability density contours have been taken from
  zero to a half of the maximum value of the probability density; in the case
  of the velocity field, contours are taken from $v=-1$ to $v=1$.
  The initial conditions for the trajectories have been taken following a constant
  distribution along the aperture for all cases to better appreciate the
  border effect on the trajectory dynamics.}
\end{figure}

So far we have commented on properties related to the construction of wave functions with
different initial shapes, which physically describe ways in which a shutter operates on an incoming
wave larger than its opening (e.g., a plane wave acted by a collimating slit).
Let us now focus on the subsequent time-evolution of such waves.
To that end a series of contourplots with the corresponding Bohmian trajectories
have been represented in Fig.~\ref{fig2} for each case considered in Table~\ref{tab1}.
The six top panels represent the time-evolution (vertical axis; normalized to the
recurrence time $\tau_r$) of the corresponding probability densities, while the six
bottom panels (labeled with a prime) describe the evolution of the associated velocity
fields.
In both cases, and for each function, a set of 20 Bohmian trajectories (white solid lines)
is also shown, with the initial conditions evenly distributed along the opening (or a bit further
away for the Gaussian wave function, just to also sample the dynamics of its ``wings'').
Perhaps such a distribution can be considered as misleading, since it is not a bona fide
representation or mapping of the evolution of the probability density $\rho(x,t)$.
However, the purpose here is not to illustrate this behavior, which can be found elsewhere
(see, for instance, \cite{sanz:JPCM:2002} for diffraction and interference in the open), but
to get a glimpse on the features characterizing diffraction under confinement conditions,
which are more prominent for marginal trajectories than from those associated with large
values of the probability density.

By inspecting the behavior of the probability density, the first we notice is the
presence of the two kind of symmetries mentioned earlier.
The space mirror symmetry displayed by the probability is very apparent for the whole evolution
of the wave function, from $t=0$ to $t=\tau_r = 397.9$, although the set of Bohmian trajectories
reveals the specificities of the dynamics, that is, there is a fast motion from some maxima to
others, while avoiding those regions where $\rho$ is negligible.
This is particularly relevant near the boundaries of the box: although initially the trajectories
spread very fast towards the boundaries, after reaching them they start undergoing a series of bounces
in order to avoid staying close.
Nonetheless, except in the case of the square function, where trajectories display fractal features
\cite{sanz:JPA:2005} and close to the borders they undergo very fast oscillations, in the other five cases
the border trajectories are relatively well-behaved, particularly in the Gaussian case.
Besides, the trajectories also make apparent that the system, for practical purposes, behaves as composed
of two independent halves, since the flux dynamics for $x<0$ does not mix with that for $x>0$, and vice versa.
This dynamical behavior is well understood if we look at the carpets corresponding to the velocity field
(bottom panels), characterized by the property of mirror antisymmetry, i.e., $v(-x,t) = - v(x,t)$.
The pronounced regions where the velocity field has large values are characterized by sudden and also large
values of the modulus of its first derivative, which provokes a fast dispersion of the trajectories.
On the other hand, the trajectories tend to accumulate in the regions where the first derivative of the velocity
field is relatively small and smooth.

When we examine the probability and velocity carpets along time, the second symmetry, namely the time-reversal
symmetry, immediately becomes apparent.
After a very fast initial boost, the wave function starts undergoing different recurrences by interference
after having interacted with the box boundaries, which generate the specific pattern of the carpet.
Now, interestingly, at $t = \tau_r/2$, there is a neat recollapse of the wave function, which gathers two features:
the probability density is split up in the form of a coherent superposition of two identical images of the initial
density, each one centered just at the center of each half of the box.
If we look at the velocity carpets, what happens is that the flux is eventually confined within these two localized
regions at $t=\tau_r/2$, that is, the trajectories are constrained to these two regions, like if there where two
openings precisely at such positions.
From this time on, the behavior of the system reverts until we observe a full recollapse of the wave to its initial state
(neglecting the global phase factor accumulated with time, which is dynamically irrelevant, as it was pointed out in
Sec.~\ref{sec2}).

%%%%%%%%%%%%%%%%%%%%%%%%%%%%%%%%%%%%%%%%%%%%%%%%%%%%%%%%%%%%%%%%%%%%%%%

\subsection{Influence of the mass: from geometric shadows to wave features}
\label{sec32}

\begin{figure}[t]
 \centering
 \includegraphics[width=\textwidth]{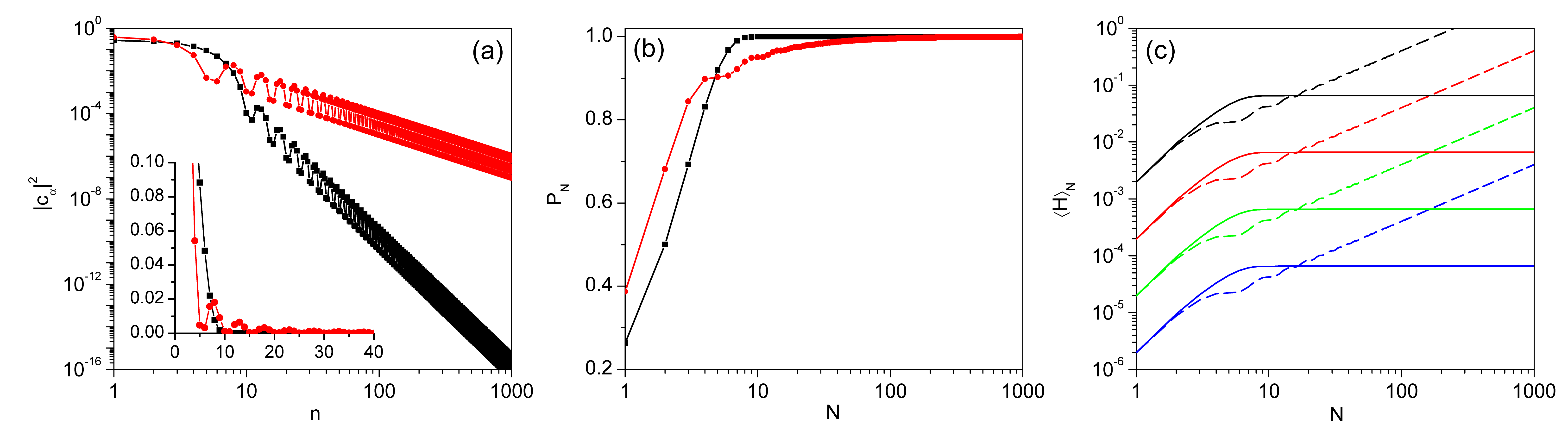}
 \caption{\label{fig3}
  (a) Weights $|c_\alpha|^2$ associated with each one of the components ($n$, with $\alpha = 2n-1$) used in the
  reconstruction  of wave functions with square (red circles) and half-cosine squared (black squares) wave
  functions (see Table~\ref{tab1}).
  In all cases: $L=50$, $w=10$ and $\hbar = 1$.
  For a better visualization, log-scale has been used in box axes; the linear-scale
  plot is displayed in the inset of the figure.
  (b) Probability $P_N$ as a function of the number $N$ of eigenfunctions for the
  two cases considered in panel (a).
  (c) Expectation value of the Hamiltonian, $\langle H \rangle_N$ as a function
  of the number $N$ of eigenfunctions for different values of the mass:
  $m = 1$ (black), $m = 10$ (red), $m = 100$ (green), and $m = 1000$ (blue).
  Solid lines represent results for the cosine-squared wave function, while dashed
  lines refer to a square wave function.}
\end{figure}

As can be seen in Table~\ref{tab1}, the system mass has no influence on the superposition itself.
In Figs.~\ref{fig3}(a) and (b) the weights $|c_\alpha|^2$ and the probability $P_N$ are shown
for the square and half-cosine squared wave functions.
We have chosen here these two particular functions, because they both are nonzero only within the interval
$|x|\leq x/2$, with the particularities that the former is an example of non-differentiable function and the latter
is close in behavior to the Gaussian.
Since the initial spectral decomposition of the wave function does not depend on the system mass, the plots
in these figures (for each function) are the same for the four masses considered in this section: $m=1$,
$m=10$, $m=100$ and $m=1000$.
Figure~\ref{fig3}(a) allows us to observe the oscillatory behavior of the weighting coefficients in
both cases, which explains the also oscillatory behavior of $P_N$ or the stepped structure of $\langle \hat{H}\rangle_N$
that we already saw in the previous section.
Interestingly here, when comparing the square and the half-cosine squared functions, we notice that while the
contribution of the eigenfunctions to the superposition (measured through $|c_\alpha|^2$) decreases slowly with $n$
for the former, the decrease is very fast for the latter (the same has also been observed for the other wave functions).
For example, while about 10 eigenfunctions have a weight above $10^{-4}$ for the half-cosine squared function,
there are about 100 eigenfunctions in the case of the square function, which explains why $\langle \hat{H}\rangle_N$
displays very clear steps in the latter case, while the same cannot be seen for the former [see Fig.~\ref{fig1}(c)].
The linear scale in the inset makes more apparent how, while the $|c_\alpha|^2$ coefficients are negligible beyond
$n=10$ for the half-cosine squared function, the same does not happen for the square function.
The manifestation of this fact can be readily seen in Fig.~\ref{fig3}(b): $P_N$ is already about 1 for $n \approx 10$
for the half-cosine squared function, while for the square function it converges very slowly to 1.
Furthermore, from a simple least square fitting, we have observed that the $|c_\alpha|^2$ decay as $n^{-2}$ for the
square function and as $n^{-6}$ for the half-cosine squared, which has interesting implications and an explanation for the
unbound increase of the expectation value of the energy in the case of the square function.
As seen in Sec.~\ref{sec2}, the eigenenergies increase with $n$ approximately like $n^2$.
So, if we compute the expectation value of the energy, we will have something like
\be
 \langle \hat{H} \rangle \propto \sum_n n^\beta n^2 ,
\ee
with $\beta$ being the exponents obtained from the fittings.
Accordingly, for the square function, we have
\be
 \langle \hat{H} \rangle \propto \sum_n 1 \to \infty ,
\ee
which is unbound, while for the half-cosine squared function we obtain a convergent series,
\be
 \langle \hat{H} \rangle \propto \sum_n \frac{1}{n^4} = \zeta (4) = \frac{\pi^2}{90} .
\ee
These are precisely the behaviors observed in Fig.~\ref{fig3}(c) for each mass.

\begin{figure}[t]
 \centering
 \includegraphics[width=0.7\textwidth]{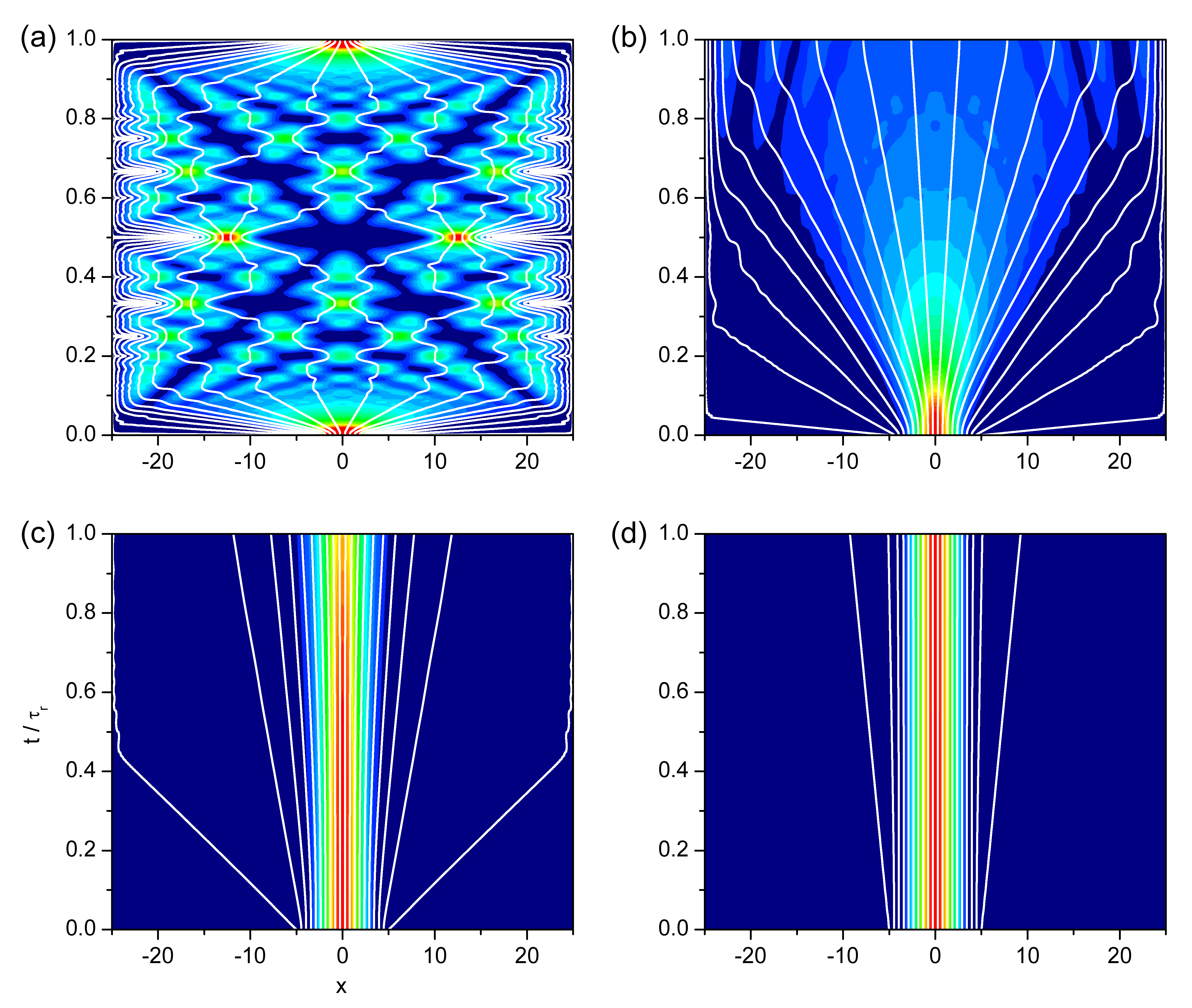}
 \caption{\label{fig4}
  Contour-plots showing the quantum carpets displayed by a half-cosine squared wave
  function (see Table~\ref{tab1}) along its evolution and for different values of
  the mass: (a) $m = 1$, (b) $m = 10$, (c) $m = 100$, and (d) $m = 1000$.
  A set of Bohmian trajectories with equidistant initial positions has also been
  superimposed in order to illustrate the dynamical evolution of the flux,
  particularly at the borders of the lattice.
  In all cases: $N=200$, $L=50$, $w=10$ and $\hbar = 1$.
  In the first panel, for a better visualization, the contours have been taken from
  zero to half the maximum value of the probability density; in all cases, the
  transition from darker (dark blue) to lighter (red) colors indicates increasing
  density values.}
%  The trajectory initial conditions have been taken following a constant
%  distribution along the aperture for (a), and a Gaussian distribution
%  for the remaining cases.}
\end{figure}

From a dynamical perspective, though, mass plays an important role in the time-evolution of the system, as
can easily be seen by inspecting the behavior of the expectation value of the energy.
This influence arises through the kinetic operator, thus here going like $m^{-1}$, as seen in Fig.~\ref{fig3}(c) for the four
masses referred to above.
Since the wave function is constituted by the same eigenfunctions and in the same proportion
(for a given initial wave function), the log-log curves for $\langle \hat{H} \rangle_N$ are always parallel, decreasing in the
same proportion in which $m$ increases.
In other words, a larger inertia implies a slower diffraction.
This effect has an interesting manifestation in the time-evolution of the wave or, equivalently, the corresponding quantum carpet.
According to (\ref{eq42}), the recurrence time $\tau_r$ increases proportionally to $m$, which means that the diffraction and
subsequent diffusion of the wave slows down.
The masses considered here increase gradually in one order of magnitude, which means that the corresponding recurrence
times are also going to increase in the same way.
Thus, if we consider as a reference the recurrence time for $m=1$, i.e, $\tau_r = 397.9$, we already notice a remarkable suppression
of the system diffusion when the mass has been increased by just one order of magnitude, as seen Fig.~\ref{fig4}(b) when compared with
Fig.~\ref{fig4}(a).
In Fig.~\ref{fig4}(b) we notice that all the structure of the quantum carpet associated with interference is completely absent; we only
observe the effect of the initial diffraction undergone by the wave function and the bounces at the boundaries of the box, apart from
some marginal interference, which becomes relevant almost at the end of the evolution.
If the mass is increased by another order of magnitude, as seen in Fig.~\ref{fig4}(c), there is still some flux associated with the edges of
wave function that can reach the boundaries of the box, but essentially all the flux remain confined within a region around the wave
function, which is slightly diffracted.
Finally, when the mass is increased by three orders of magnitude above the reference mass, the wave function does not show much
diffraction, as it is shown in Fig.~\ref{fig4}(d).
In this latter case, notice that the trajectories remain nearly parallel one another.

It is worth noting that this latter case is the quantum analog to the geometric optics limit, where diffraction effects are neglected
behind an opening.
Consider the shutter is illuminated by monochromatic light with a negligible wavelength compared to the shutter width ($\lambda \ll w$),
and that a screen is allocated a certain distance $L = \tau_r/c$.
In a first approximation, the imaging problem at $L$ can be described by means of the geometric optics.
Accordingly, there will be a spot of light just in front of the shutter, with nearly its same width (if the incident radiation is a plane wave), and
shadow everywhere else.
However, if the distance to the screen increases, a series of diffraction traits start appearing because light start displaying its wave behavior.
Furthermore, if a constraint is imposed on its spatial diffusion (reflecting walls, e.g., mirrors) and $L$ becomes larger and larger, interference traits
will manifest and eventually we will observe the same behavior as in our case for $m=1$ (or a periodic representation of the same if the length
of the box or, equivalently, the propagation time is further increased).
With the matter wave we have exactly the same, as seen in Fig.~\ref{fig4}, if the wave entering the cavity is highly coherent.
For very short times or very large masses, the system can be described in a first approximation with classical mechanics, since Bohmian trajectories
are going to closely behave as Newtonian ones.
Actually, this situation is what can be denoted as the Ehrenfest-Huygens regime \cite{sanz:AJP:2012}.
However, as time increases or smaller masses are considered, wave-like features, like diffraction or interference, become dominant in the evolution displayed
by the trajectories and classical mechanics is no longer a good description of the system dynamics --- notice in Fig.~\ref{fig4}(d) that, if the system is left
to evolve up to the corresponding recurrence time (a thousand times larger), we shall observe a picture exactly the same as in Fig.~\ref{fig4}(a).
Typically, according to the standard view, wave and particle behaviors are incompatible.
This simple example here shows that this statement is not true, but that all depends on the scale of time (or mass) considered to analyze
the system.
Within this context, classical mechanics (or geometric optics, if we are dealing with light) is just a first-order approximation to the behavior displayed
by the system in the very short term, regardless of its mass --- of course, another matter beyond the scope of this discussion, but also very important at
a fundamental level, is the whether by more mass one means more complex, i.e., a many-body object.

%%%%%%%%%%%%%%%%%%%%%%%%%%%%%%%%%%%%%%%%%%%%%%%%%%%%%%%%%%%%%%%%%%%%%%%%

\subsection{Influence of the confining boundaries: interference (patterning) structure}
\label{sec33}

\begin{figure}[t]
 \centering
 \includegraphics[width=\textwidth]{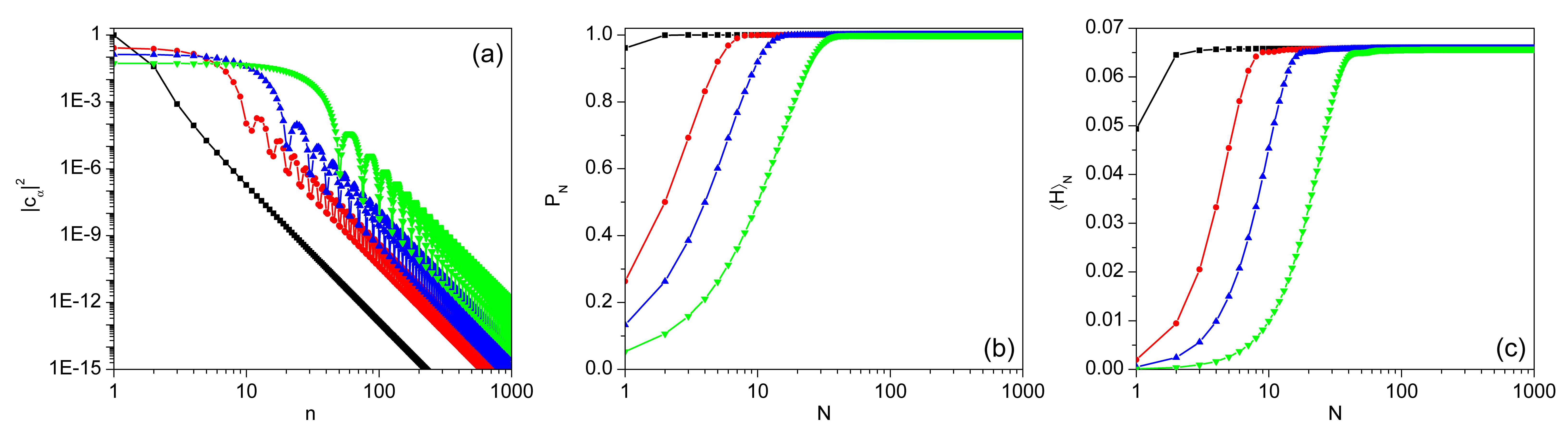}
 \caption{\label{fig5}
  (a) Weights $|c_\alpha|^2$ associated with each one of the components ($n$, with $\alpha = 2n-1$) used in the
  reconstruction of a half-cosine squared inside a box with different lengths: $L = 10$ (black squares)
  $L = 50$ (red circles), $L = 100$ (blue triangles), and $L = 200$ (red diamonds).
  For a better visualization, log-log scale has been used in both axes.
  (b) Probability $P_N$ as a function of the number $N$ of eigenfunctions for the cases
  cases considered in panel (a).
  (c) Expectation value of the Hamiltonian, $\langle H \rangle_N$, as a function
  of the number $N$ of eigenfunctions.
  In all cases, the shutter width is $w = 10$ and the system mass $m = 1$ (with $\hbar = 1$).}
\end{figure}

To complete the analysis, the effects of the size of the box on the system have also been studied, since the influence of this parameter
not only comes from the recurrence time (\ref{eq42}) (as $L^2$), but also through the momenta $k_\alpha$, according to (\ref{eqkalp})
(as $L^{-1}$), and the relative weight $|c_\alpha|^2$ in the superposition (as $L^{-1}$).
As in the previous sections, before going into detail with the dynamics, let us get some conclusions from the corresponding superpositions.
To that end, using again a half-cosine squared wave function, four different box sizes have been considered, including as a reference the
previous one $L=50$.
These sizes range from $L=w$, so that the shutter opening covers the whole box, to $L = 20 w$, which is large enough compare to the
shutter opening.
As can readily be seen in Fig.~\ref{fig5}(a), as $L$ increases the superposition becomes more and more structured, including a larger
number of eigenfunctions with the same relative weight.
For example, if we consider a threshold of contributions around 0.1\% or above, for $L=w$ we have one major contribution, which is
nearly a 100\% of the superposition, the next falls below 10\%, and the third drops to about 0.1\%.
For $L=5w$, there are 9 contributions above 0.1\%, with 3-4 of them close in importance, above 10\%.
These numbers remarkably increase with $L=10w$ and $L=20w$, as seen in the figure.
Actually, the trend is that the number of eigenfunctions contributing with nearly the same weight increases with $L$.
If we consider those contributions which differ from the first one up to 25\%, i.e.,
\be
 \Delta_{1,\alpha} = \left( 1 - \frac{ |c_\alpha|^2}{|c_1|^2} \right) \times 100\% ,
\ee
we find that for $L=w$ there is only 1, for $L=5w$ there are 2, for $L=10w$ there are 4, and for $L=20w$ there are 17, which follows
a nearly quadratic dependence on $L$.
A similar trend can also be seen in case of $P_N$ and $\langle \hat{H} \rangle_N$, as it is shown in panels (b) and (c) of Fig.~\ref{fig5},
where an increasing number of $L$ means that a remarkable number of eigenfunctions is needed to better represent the original wave
function and therefore an slower convergence to it and its energy.

\begin{figure}[t]
 \centering
 \includegraphics[width=\textwidth]{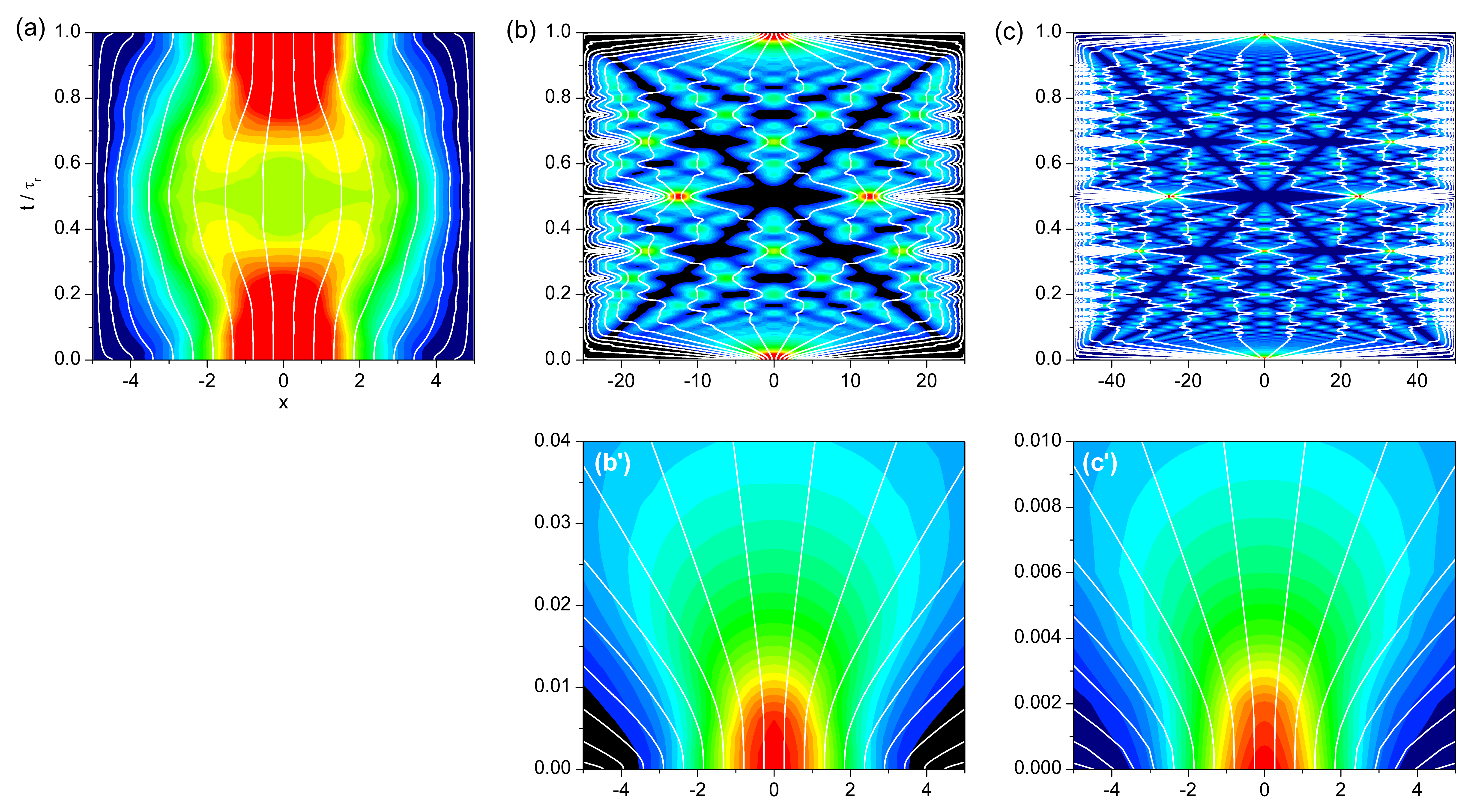}
 \caption{\label{fig6}
  Contour-plots showing the quantum carpets displayed by a cosine-squared wave
  function (see Table~\ref{tab1}) along its evolution and for different values of
  the box size: (a) $L = w$, (b) $L = 5w$ and (c) $L = 10w$.
  Panels (b') and (c') are enlargements of the regions of (b) and (c), respectively,
  for the same time displayed in panel (a).
  A set of Bohmian trajectories with equidistant initial positions has also been
  superimposed in order to illustrate the dynamical evolution of the flux,
  particularly at the borders of the lattice.
  In all simulations here: $N=200$, $w=10$ and $m=1$ (with $\hbar = 1$).
  In the first panel, for a better visualization, the contours have been taken from
  zero to half the maximum value of the probability density; in all cases, the
  transition from darker (dark blue) to lighter (red) colors indicates increasing
  density values.}
%  The trajectory initial conditions have been taken following a constant
%  distribution along the aperture for (a), and a Gaussian distribution
%  for the remaining cases.}
\end{figure}

If we now go to the corresponding quantum carpets, displayed in Fig.~\ref{fig6} for $L=w$, $L=5w$ and $L=10w$,
we notice an increasing degree of complexity and structuring with increasing $L$, which is expected as the number of
eigenfunctions involved, and hence the number of frequencies $\omega_{\alpha,\alpha'}$, also increases.
This gives rise to a highly noisy dynamics, as seen through the corresponding Bohmian trajectories, which comes from the fact
that interference traits become more prominent due to the appearance of more profiled dips and ridges [compare panels (b)
and (c)], which forces the trajectories to jump relatively fast from some regions to others, since the velocity field is too large
in between.
Nonetheless, for short times, of the order of 1/25 of the $\tau_r$ corresponding to the case of $L=5w$ and 1/100 of the one
corresponding to $L=10w$, we find a very similar early-time evolution, as seen in panels (b') and (c'), respectively,
which is in correspondence with the fact that at these
stages there is not time enough yet to notice the fine structuring effect coming from all the main contributions (many more in
the latter case than in the former, as seen in the upper panels).

%%%%%%%%%%%%%%%%%%%%%%%%%%%%%%%%%%%%%%%%%%%%%%%%%%%%%%%%%%%%%%%%%%%%%%%
%%%%%%%%%%%%%%%%%%%%%%%%%%%%%%%%%%%%%%%%%%%%%%%%%%%%%%%%%%%%%%%%%%%%%%%

\section{Concluding remarks}
\label{sec4}

From a dynamical viewpoint, we have that the delocalization of a released matter wave is analogous to the diffraction it
undergoes after crossing an opening --- in this latter regard, the opening would act as the localizing element and its
subsequent crossing would play the role of the release.
On the other hand, regardless of the initial physical context considered (whether a trapped atomic cloud or a diffracted
atomic or molecular beam), if some extra boundaries are added, the new confining conditions will produce the appearance
with time of a series or recurrences.
The pattern that develops with time is commonly known as a quantum carpet, which displays some symmetries in both
space and time according to the interference of the wave with the new confining boundaries.
Actually, at some time, a full revival of the initial state (except for a global phase factor) is observed, which is repeated in
time once and again unless some dissipative or decohering mechanisms act on the system.
This is particularly remarkable in the case of the well-known problem of the particle in a one-dimensional box, assuming
such a particle is nonrelativistic, spinless and with mass~$m$.

In this work we have focused on this classical problem with the purpose to determine which are the main elements that
affect the evolution of the bound diffraction process, and more specifically how such elements influence the symmetry
displayed by the wave function and its associated flux along their evolution.
To that end, we have combined the standard spectral decomposition of the initially localized state in terms of
coherent superposition of energy eigenstates with a Bohmian description of its eventual dynamics.
Indeed, the possibility to decompose the initial state in this way has been profitably used to devise a simple and efficient
analytical algorithm that makes easier and more accurate the computation of velocity fields (flows) and trajectories, since
the value of the associated velocity field can be exactly obtained at each position of the configuration space.

As it has been shown, these two tools (spectral decomposition and Bohmian trajectories) constitute two rather suitable
tools to explore and analyze the problem of the formation of space-time patters inside the cavity in terms of the three
key elements that  rule the bound diffraction process and the consequent matter-wave dynamics: the shape of the initial
wave function, the mass of the particle considered, and the relative extension of the initial state with respect to the total
length spanned by the cavity.
Specifically, from the spectral decomposition we have been able to identify how each one of these elements
contributes to the superposition that generates the corresponding localized matter wave as well as to its eventual
evolution; the Bohmian analysis, on the other hand, reveals aspects connected to the diffraction dynamics and the
subsequently developed interference traits, such as the origin of the characteristic symmetries displayed by these
systems or the appearance of recurrences and full revivals of the initial state.
Furthermore, we have also observed that, because of the presence of confining boundaries, even in the
cases of an increasingly large box length, no Fraunhofer-like diffraction features can ever be observed at any time,
as it is the case of the analogous unconstrained waves.
This is because of the relatively fast development of a phase field spreading through the whole of the box, which
becomes faster as the box length becomes larger and larger.

From the analysis presented, and particularly if we take into account the explicit expression for
the recurrence time, Eq.~(\ref{eq42}), it may seem that from the three elements involved in development
of quantum carpets, the mass could have been neglected, since in principle it only plays the role of a
scaling factor.
To some extent, if we consider the quantum carpet as a whole, this is true, which can be taken as a practical
advantage in the design of wave-function propagation algorithms based on the above mentioned spectral
decompositions.
However, when we think in terms of time as a limiting parameter, which may allow or not to observe a full
quantum carpet, this conception changes, for it readily brings out the question of the classical limit in
a rather challenging way: although quantum systems are commonly associated with small masses (or, equivalently,
small objects), this simple model shows that classicality with larger masses is a sort of illusions, because
if we wait long enough, quantum features eventually appear.
Therefore, we can see that classical behaviors are not only determined by the size of the object, as it is
commonly taught in standard quantum mechanics courses, because in such a case it is only a matter of waiting
long enough to observe quantum behaviors --- it is not just a matter of size, but complexity and, therefore,
the influence of other phenomena, such as decoherence by entanglement with environmental degrees of freedom
(regardless of whichever system the environment represents).
The time the wave function remains with its nearly unchanged shape is what we typically regard as the Ehrenfest
time when referring to wave packets with negligible spreading \cite{sanz:AJP:2012} --- this is the Huygens regime
in wave optics, previous to the Fresnel one, when nearly well-defined shadows can still be seen before oscillatory
behaviors start appearing.

The analysis here has been applied to matter waves.
However, we would like to highlight that both the methodology and conclusions are also valid in the case of light
propagation through optical fibers, where the input (light) state can be constructed by just selecting the appropriate
electromagnetic modes.
Actually, notice that under paraxial conditions, Helmholtz's equation, which describes the distribution of electromagnetic
energy inside the fiber, acquires the form of a Schr\"odinger-like equation \cite{sanz:JOSAA:2012}.
This thus opens an alternative procedure to develop efficient Bohmian-based numerical methodologies to explore and
control the dynamics of bound quantum and optical systems in a rather simple fashion.
Furthermore, a rather direct extension of this work seems to be re-examining the so-called diffraction in time phenomenon
\cite{moshinsky:PhysRev:1952,moshinsky:AJP:1976,schuch:JPA:2001,muga:JPA:2006,muga:PhysRep:2009,horwitz:FoundPhys:2007}
under (time) confinement conditions \cite{muga:JPA:2006}.

%%%%%%%%%%%%%%%%%%%%%%%%%%%%%%%%%%%%%%%%%%%%%%%%%%%%%%%%%%%%%%%%%%%%%%%

\ack

Financial support from the Spanish MINECO (Grant No.\ FIS2016-76110-P)
is acknowledged.

%%%%%%%%%%%%%%%%%%%%%%%%%%%%%%%%%%%%%%%%%%%%%%%%%%%%%%%%%%%%%%%%%%%%%%%

\section*{References}

%\bibliography{references}

\providecommand{\newblock}{}

\end{document}